\documentclass[proof]{WileyASNA-v1}

\articletype{Article Type}%
\usepackage{graphicx,epsfig}
\usepackage{tabularx,tabulary,threeparttable,float,lscape}
\usepackage{url}
\newcommand{\specialcell}[2][c]{%
  \begin{tabular}[#1]{@{}c@{}}#2\end{tabular}}

\received{February 6, 2020}
\revised{June 26, 2020}
\accepted{XXXX, 2020}

\begin{document}

\title{Isolated AGNs NGC 5347, ESO 438-009, MCG-02-04-090, and J11366-6002: Swift and NuSTAR joined view\protect\thanks{Swift/XRT, Swift/BAT, NuSTAR}}

\author[1]{Vasylenko A.A.}

\author[1]{Vavilova I.B.*}

\author[1]{Pulatova N.G.}

\authormark{Vasylenko A.A. \textsc{et al}}

\address[1]{\orgdiv{Department for Extragalactic Astronomy and AstroInformatics}, \orgname{Main Astronomical Observatory of the National Academy of Sciences of Ukraine}, \orgaddress{\state{27, Akademik Zabolotny St., 03143, Kyiv}, \country{Ukraine}}}

\corres{*Vavilova Irina. \email{irivav@mao.kiev.ua}}

\presentaddress{Main Astronomical Observatory, NAS of Ukraine, 27, Akademik Zabolotny St., 03143, Kyiv, Ukraine}

\abstract{We present the spectral analysis with the Nuclear Spectroscopic Telescope Array (NuSTAR) of four isolated galaxies with active galactic nuclei selected from 2MIG catalogue: NGC 5347, ESO 438-009, MCG-02-09-040, and IGR J11366-6002. We also used the Swift/Burst Alert Telescope (BAT) data up to $\sim$ 150 keV for MCG 02-09-040, ESO 438-009, and IGR J11366-6002 as well as the Swift/XRT data in 0.3--10 keV energy band for NGC 5347, ESO 438-009, and IGR J11366-6002.

All the sources appear to have the reflected spectrum component with different reflection fractions in addition to the primary power-law continuum. We found that power-law indices for these sources lie between 1.6 to 1.8. The spectra of two sources, NGC 5347 and MCG-02-09-040, show the Fe $K_{\alpha}$ emission line. For both of these sources, the Fe $K_{\alpha}$ lines have a large value of EW $\sim$ 1 keV. The X-ray spectrum of NGC 5347 is being best fitted by a pure reflection model with $E_{cut} \sim 117$ keV and without the presence of any additional primary power-law component. 
We also found that X-ray spectrum of MCG -02-09-040 shows the presence of heavy neutral obscuration of $N_{H}\sim10^{24}~cm^{-2}$. However, this provides a non-physical value of reflection in the case with fitting by a simple reflection model. A more appropriate fit is obtained with adopting physical Monte Carlo based model as BNTorus. It allowed us to determine the absorption value of \textbf{$N_{H}\sim 1.04\times 10^{24}~cm^{-2}$} and reasonable power-law index of $\Gamma \approx 1.63$. Results for MCG -02-09-040 are presented for the first time.
}

\keywords{galaxies, active galactic nuclei, X-ray spectra; objects: NGC 5347, ESO 438-009, MCG-02-09-040, IGR J11366-6002}

\maketitle

\footnotetext{\textbf{Abbreviations:} AGN, active galactic nuclei; NuSTAR, Nuclear Spectroscopic Telescope Array space mission; BAT, Swift/Burst Alert Telescope}

\section{Introduction}\label{sec1}
A deep systematic analysis of detailed X-ray spectral parameters of active galactic nuclei (AGN) in the isolated galaxies, as a specific class of galaxies, has been not presented yet. Some general properties of nuclear activity and X-ray spectra of isolated AGNs from the 2MIG catalogue (\cite{Karachentseva2010}) based on the 2MASS (\citet{Skrutskie2006}) were discussed by ~\citet{Hernandez2014}, \citet{Pulatova2013, Pulatova2015}, \citet{Vavilova2015a} and others. 

Hern\'{a}ndez-Ibarra et al. (2013) using two samples of isolated galaxies, have found that the incidence of non-thermal nuclear activity is presented in 31 \% and 40 \% of isolated galaxies with strong emission lines in these samples. The authors concluded that the internal secular evolution in such isolated galaxies could be considered responsible for nuclear activity. They also found the absence of Type 1 AGN among broad-line active galaxies. This leads to the conclusion that the broad-line region can be formed only in the higher accretion rates/luminosities environment triggered by an interaction mechanism. The recent mergers or interactions as a cause of nuclear activity in the isolated AGNs are of a prevalent point of view (see, for example, \cite{Alonso2007}. 

\cite{Pulatova2015} studied the general wide-band properties of isolated AGNs at z\,$\leq$ 0.1. It was obtained that most of selected 2MIG AGNs are faint X-ray sources (excluding IC2227 and UGC 10120) with a mean flux $9.4\times 10^{-12}~erg~cm^{-2}~s^{-1}$ in 2--10 keV range for all the 2MIG isolated AGNs without faint nearby companions: a mean luminosity is $5.7\times 10^{42}~erg~s^{-1}$. 
This conclusion corresponds well to the results by \cite{Vavilova2015a}, who presented a spectral analysis of four isolated AGNs (NGC 1050 (Sy2 type), ESO 317-038, ESO 438-009 (Sy1.5 type), NGC 2989) using XMM-Newton and Swift/XRT/BAT data.  

\cite{Risaliti1999} have characterized NGC 5347 as a Compton thick AGN with EW(Fe)>1.9 keV. It was observed by BeppoSAX in 2--200 keV band that allowed \cite{Dadina2008} to determine the spectral shape of this AGN for disentangling Compton emission and reflection spectra components. \cite{LaMassa2012} analyzed NGC 5347 using the Chandra data and found that spectrum in 2--10 keV shows a quite flat photon index $\Gamma$=1.53 as well as a moderate absorption of $N_{H}\sim 33.0\times 10^{22}~cm^{-2}$.

\cite{Koss2010} have investigated Swift/BAT hard X-ray AGN sample and revealed a significant increase in the number of hard X-ray AGNs in the merging galaxies. Besides that, such galaxies are also characterized by a large value of the hard X-ray to [OIII] flux ratio, which can be used to study the X-ray properties of isolated AGNs.

Due to a high sensitivity and better angular resolution in hard X-rays (i.e. >10 keV) as well as capability to focus photons at that energies, the NuSTAR allows us to understand better physical properties of the isolated galaxies with AGNs. In this work we study the X-ray spectra of four isolated AGNs (NGC 5347, MCG-02-09-040, ESO 438-009, IGR J11366-6002) using NuSTAR data in 3--79 keV and Swift/XRT/BAT data (see, Table~\ref{tab1}). In Section 2 we describe the data reduction procedures applied for processing of the NuSTAR and Swift X-ray observations. In Section 3 we present a time-averaged spectral study of these galaxies with their best-fit models and values of spectral parameters. Discussion and concluding remarks are given in Section 4.
In Appendix A we discuss the time-resolved spectral analysis of the IGC J11366-6002 using NuSTAR data.

\section{Observations and data reduction}\label{sec2}
\subsection{The NuSTAR data reduction}
The NuSTAR data reduction was performed with the NuSTARDAS 1.6.0, which is included in the HEASOFT package 6.19, using calibration files from NuSTAR CALDB v20160502. The data were processed with standard settings by \textit{nupipeline} task for each of FPMA and FPMB detectors. Source counts were extracted in circular regions with radius 40--50 arcsec. The background counts were extracted from the region between two circles, centered on the source position, with the inner radius of 80--90 arcsec and outer of 120--130 arcsec. Spectra and light curves were processed with the standard \textit{nuproducts} tool.

Due to the low-quality data of NGC 5347 and MCG-02-09-040 in addition to the standard ''science'' data (or data from mode 1) we included also the ''spacecraft science'' (mode 6) data following the procedure outlined by \cite{Walton2016} to maximize the signal-to-noise. This increased a total good exposure time by $\sim$ 5 ks for the first one and $\sim$ 2 ks for the second one. The FPMA/FPMB spectra were limited to the 3.0--50 keV range because of high background at E > 50 keV. The obtained FPMA and FPMB spectra were fitted jointly, but without merging them into single spectrum. Each of spectra was grouped to a minimum of 25 counts per bin.

\subsection{The Swift data reduction}
All the studied sources have X-ray observations from XRT instrument on board the Swift satellite (see, Table~\ref{tab1}). The XRT data reduction was performed with \textit{xrtpipeline} task, which is included in the HEASOFT package 6.19. All the data were extracted only in the photon counting (PC) mode adopting the standard grade filtering 0--12 for PC. Source counts were extracted in a circular region with a radius of 20 arcsec. The background counts were extracted using \textit{xselect} from the region between two circles centered on the source position with the inner radius of 30 arcsec and outer of 50 arcsec. We created appropriate ancillary response files using \textit{xrtmkarf}. For increasing the spectral quality we combined the obtained XRT spectra into one for each source using \textit{addspec v1.2.1} tool. We also added the corresponding background and the ARFs were combined into single using \textit{addarf 1.2.6} tool. We did not use observation data for MCG -02-09-040 by Swift/XRT due to poor quality.

To improve our analysis for ESO 438-009 and IGR J11366-6002 we used also the Swift/BAT stacked spectra at energies 14--195 keV from the 70-month Hard Survey catalog\footnote{\url{https://swift.gsfc.nasa.gov/results/bs70mon/}} (\cite{Baumgartner2013}), which spanned 2004--2010 years. For MCG-02-09-040 we used Swift/BAT stacked spectra in 15-150 keV band from the publicly available Third Palermo Swift/BAT hard X-ray catalogue.

\section{Spectral analysis}\label{sec3}
\subsection{General notes}
The spectral analysis was performed with a spectral fitting software XSPEC v.12.8.2q. Each spectrum, except for NGC 5347, was fitted used $\chi^{2}$ statistic. In the case of NGC 5347, we apply C-statistic due to low number counts in Swift/XRT data. The quoted errors of best fit parameter values are at the 90 \% confidence level ($\Delta \chi^2=2.71$).

An additional cross-calibration constant was included to vary normalizations between each of instrument as a free parameter. These normalization offsets are found to be around 5\% as expected from instruments calibration \citep{Madsen2015}. The absorption model \textit{tbabs} \citep{Wilms2000} is used to account for the Galactic absorption. The values of these absorption were taken from Leiden/Argentine/Bonn Survey of Galactic HI\footnote{\url{https://heasarc.gsfc.nasa.gov/cgi-bin/Tools/w3nh/w3nh.pl}} \citep{Kalberla2005}. The measured galactic inclination angles were taken from HyperLEDA database. The standard cosmological parameters were adopted as follows: $H_{0}$ = 70 km s$^{-1}$ Mpc$^{-1}$, $\Omega_{\Lambda}$= 0.73, $\Omega_{M}$=0.27.

The spectral analysis for all objects has been started from application of a simple power-law model with galactic $N_{H}^{Gal}$. Then, in order to take into account the possible presence of additional neutral absorption, cut-off, reflection component or soft excess, we applied the following XSPEC models (in dependence of the studied source): \textit{ztbabs}, cut-off power-law, \textit{plcabs} \citep{Yaqoob1997}, \textit{pexrav} \citep{Magdziarz1995}, \textit{pexmon} \citep{Nandra2007} or \textit{xillver} \citep{Garcia2010, Garcia2013} model.

\begin{figure*}[t]
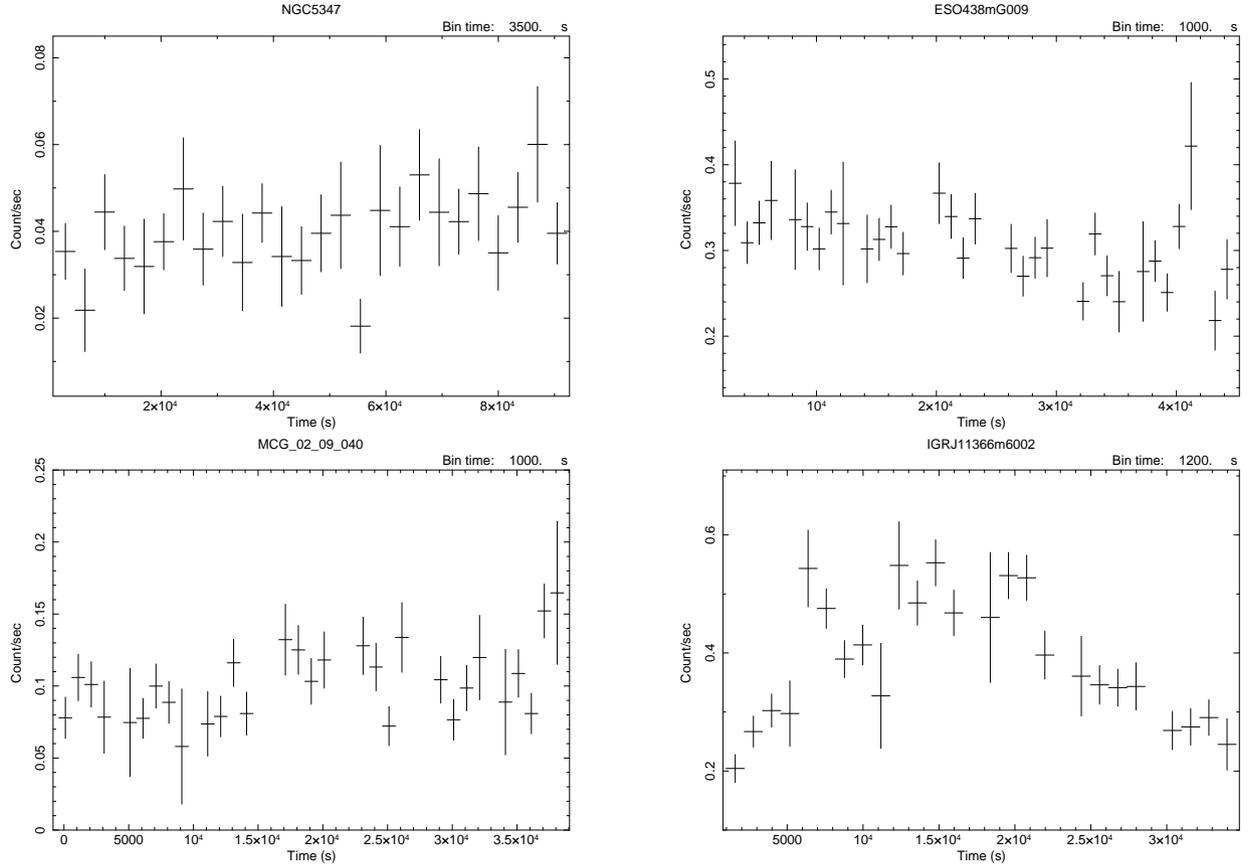

\centering
\begin{minipage}[h]{0.45\linewidth}
\epsfig{file=ngc5347_light_curve_2.eps,width=0.71\linewidth,angle=-90}
\end{minipage}
\hskip+7mm
\begin{minipage}[h]{0.45\linewidth}
\epsfig{file=eso438009_light_curve_2.eps,width=0.71\linewidth,angle=-90}
\end{minipage}
\vfill
\begin{minipage}[h]{0.45\linewidth}
\epsfig{file=mcg0209040_light_curve_2.eps,width=0.71\linewidth,angle=-90}
\end{minipage}
\hskip+7mm
\begin{minipage}[h]{0.45\linewidth}
\epsfig{file=j11366_light_curve_2.eps,width=0.71\linewidth,angle=-90}
\end{minipage}
    \caption{The background-subtracted FPMA+FPMB light curves of the isolated galaxies with AGNs in 3-79 keV: NGC 5347, ESO 438-009, MCG-02-09-040, and IGR J11366-6002. \label{fig1}}
\end{figure*}

The background-subtracted FPMA+FPMB light curves of the studied AGNs in 3-79 keV range are shown in Fig.~\ref{fig1}. The best-fit spectra of these sources are in Fig.~\ref{fig2}.

\subsection{Notes on individual sources}
\textbf{NGC 5347}. As a first fit to the NuSTAR spectra we applied a simple cut-off power-law plus gaussian line model \textit{cutoffpl+zga}. This model provides a good fit (C-stat=57/47) but with unrealistic very flat index $\Gamma = 0.03_{-0.20}^{+0.19}$ mostly due to curvature at energy above 10 keV. After that we fixed the spectral index at typical value for Seyfert galaxies $\Gamma\approx$ 1.7 (see, e.g., \cite{Dadina2008}), but this gives us a worse fit with C-stat=263/48. A much better and reasonable fit (C-stat=54/48) was obtained with \textit{pexmon} reflection model, which self-consistently includes emission Fe $K_{\alpha}$ line. The photon index becomes $\Gamma$=1.68 $\pm$ 0.14 and $E_{cut} > 38$ keV with the assumption of pure reflected spectrum, i.e. when reflection parameter is fixed $R \equiv$ -1. And even when $R$ varies freely, it leads that value of $R$ is completely unconstrained. To obtain the parameters of iron emission line, we used the \textit{pexrav+zgauss} model (C-stat=50/45). The width of the line was fixed to $\sigma$ =10 eV. Thus, the Fe K$_{\alpha}$ line energy is found to be $E_{line}$=6.46 $\pm$ 0.08 keV and the equivalent width is large $EW = 1.62 \pm 0.39$ keV as expected for a reflection-dominated spectrum.

The combined NuSTAR and Swift/XRT spectrum was fitted (C-stat=55/52) by \textit{pexmon} model in the pure reflection regime with a \textit{zbbody} component for describing the soft emission ($kT_{e} = 0.77_{-0.18}^{+0.17}$ keV). The spectral index is $\Gamma = 1.60_{-0.27}^{+0.37}$ and cut-off energy is constrained to be $E_{cut} > 63$ keV. 
The unabsorbed (i.e., absorption-corrected) luminosity are L$_{2-10~keV}=2.89_{-0.19}^{+0.22} \times~10^{41}$~erg~s$^{-1}$ and L$_{10-40~keV}=5.93_{-0.14}^{+0.16} \times 10^{41}$ erg~s$^{-1}$ in 2--10 keV and 10--40 keV ranges, respectively.

\cite{Risaliti1999} found for the NGC 5347 that EW(Fe)>1.9 keV and F$_{2-10keV}=3.8\times 10^{-13}$~erg~cm$^{-2}$~s$^{-1}$. Because of that paper does not contain information on the input data and its processing technique, we can say only that this value is a little bigger than our estimations, EW(Fe) = $1.62 \pm 0.39$ keV, but it is in agreement within margin of errors EW(Fe)=1.23…2.01). We note also that opposite to the earlier result obtained by \cite{LaMassa2012} with the X-ray data from Chandra, we were able to obtain a good fit to X-ray spectrum without using addition neutral absorption. However, it should be noted that quality of data in the soft band (<3 keV) are insufficient for clear determination of this possible absorption feature.

\begin{figure*}[t]
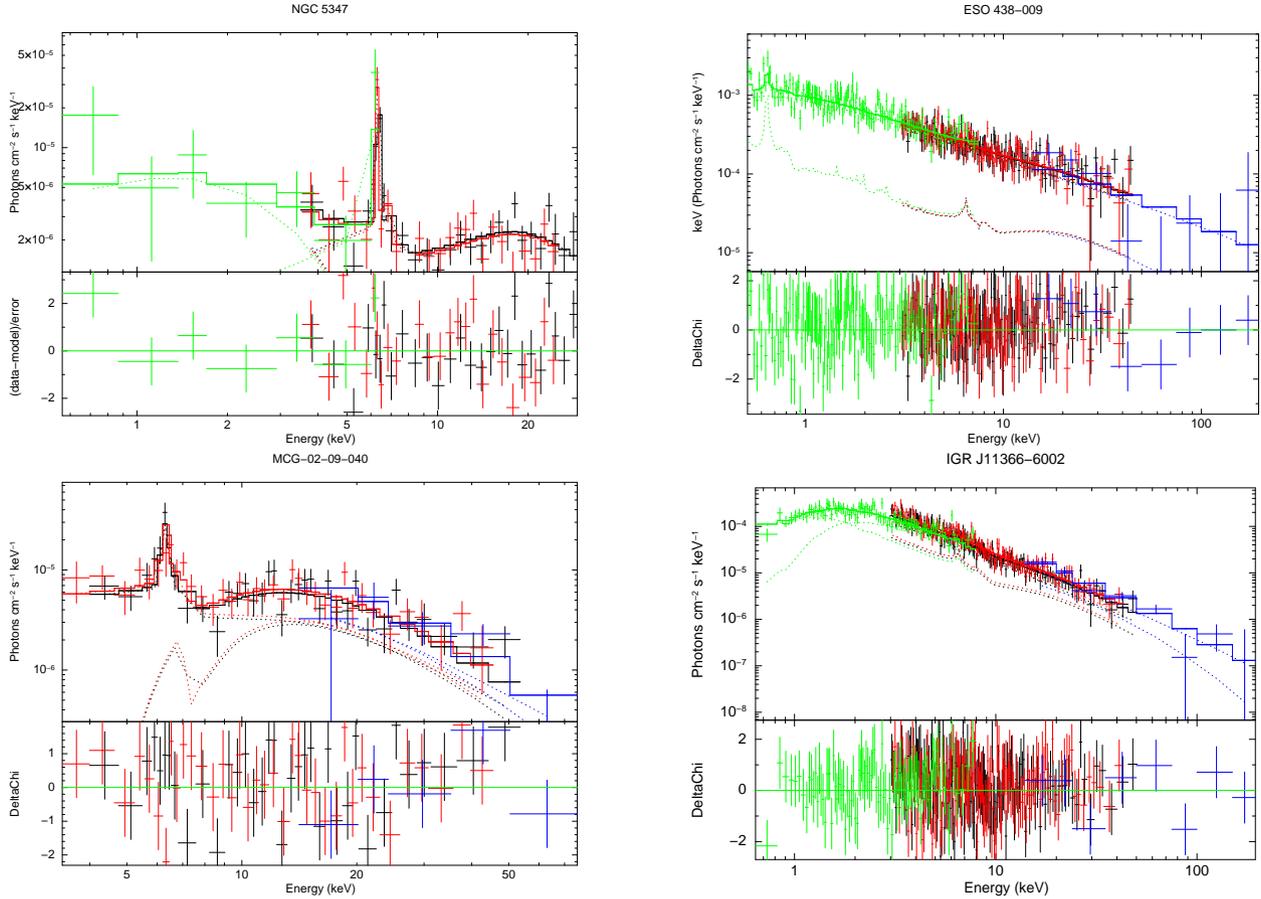

\centering
\begin{minipage}[h]{0.46\linewidth}
\epsfig{file=NGC5347_fixed_col.eps,width=0.72\linewidth,angle=-90}
\end{minipage}
\hskip+7mm
\begin{minipage}[h]{0.46\linewidth}
\epsfig{file=eso438009_spectrum.eps,width=0.72\linewidth,angle=-90}
\end{minipage}
\vfill
\begin{minipage}[h]{0.46\linewidth}
\epsfig{file=MCG0209040_fixed_col.eps,width=0.72\linewidth,angle=-90}
\end{minipage}
\hskip+7mm
\begin{minipage}[h]{0.46\linewidth}
\epsfig{file=j11366_spectrum.eps,width=0.72\linewidth,angle=-90}
\end{minipage}
    \caption{ NuSTAR plus Swift/XRT+BAT best-fit unfolded spectra with corresponding residuals for the analyzed sources. The red and black data points correspond to NuSTAR FPMA and FPMB data, respectively, the green data points are Swift/XRT data, and the blue data points are Swift/BAT data. \label{fig2}}
\end{figure*}

The best-fit spectrum with corresponding residuals for AGN in the isolated NGC 5347 is shown in Fig.~\ref{fig2}, upper-left panel; see also Table~\ref{tab2}. 

\textbf{MCG -02-09-040}. The combined NuSTAR and Swift/BAT spectrum for this AGN shows a good fit ($\chi^{2}$/d.o.f.=72/75) with a cut-off power-law absorbed by a partially covering medium. The column density found for the partial absorber is $N_{H}=112_{-16}^{+20} \times 10^{22}$~cm$^{-2}$, and the covering fraction is 0.86$\pm$0.03. The cut-off energy is constrained to be 
$E_{cut} > 108$ keV and power-law index is $\Gamma = 1.53_{-0.11}^{+0.12}$. An addition narrow Fe K$_{\alpha}$ line is detected at $E_{line}=6.45_{-0.12}^{+0.10}$ keV with an equivalent width of $EW=898_{-367}^{+351}$ eV. Such $EW$ of iron line together with the quite flat power law indicate strongly a highly absorbed AGN.

Afterwards, we fitted combined spectrum with \textit{plcabs} model \citep{Yaqoob1997} instead of the absorbed power-law model. This model describes X-ray transmission of an isotropic source of photons at the centre of a uniform sphere of matter and takes the Compton scattering into account. Employing \textit{plcabs} model together with \textit{pexmon} model (which applied in a pure reflection regime) provides an excellent fit $\chi^{2}$/d.o.f.=77/75 and gives a spectral index of $\Gamma = 2.10 \pm 0.20$. The resulting absorption is $N_{H}=186_{-40}^{+45} \times 10^{22}$~cm$^{-2}$ that together with large value of Fe $K_{\alpha}$ line equivalent width indicate that MCG -02-09-040 has the Compton-thick AGN.

We accentuate that \textit{plcabs} model includes a simple approach of Compton scattering. Namely, this model assumes a constant Compton scattering cross section equal to the Thomson cross section. As pointed out by \cite{Murphy2009}, the \textit{plcabs} model in tandem with the \textit{pexrav}-based reflection model may produce a bias of towards by overestimation of the direct continuum.
	
Thus, we apply another, more physically self-consistent models, based on Monte-Carlo simulation, such as \textit{BNTorus} and \textit{Sphere} models \citep{Brightman2011, Vasylenko2018b}. These models take into account photoelectric absorption, Compton scattering and fluorescence iron emission lines from a toroidal or spherical smooth obscured structure centered on the AGN, respectively. \textit{Sphere} model returns quite flat power-law index $\Gamma = 1.26 \pm 0.13$ and, unexpectedly, a low column density of $N_{H}=35_{-9}^{+18} \times~10^{22}$~cm$^{-2}$ with a fit statistic of $\chi^{2}$/d.o.f.=97/73. \textit{BNTorus} model, by contrast, returns a much better fit $\chi^{2}$/d.o.f.=77/76 with $\Gamma = 1.63 \pm 0.11$ and $N_{H}=104_{-21}^{+26} \times~10^{22}$~cm$^{-2}$ (see Table~\ref{tab3}). We note that $N_{H}$ obtained from latter model is slightly lower than that obtained from phenomenological model. The opening angle of the torus measured by \textit{BNTorus} is $\theta_{oa}=37_{-peg}^{+7}$ degrees, which corresponds to a covering factor of $\leq$~0.79. The inclination angle of the torus is estimated to be $\theta_{incl}=41_{-8}^{+26}$ degree.

The unabsorbed luminosities are L$_{2-10~keV}=1.43_{-0.01}^{+0.03} \times 10^{42}$~erg~s$^{-1}$ and L$_{10-40~keV}=2.17_{-0.01}^{+0.12} \times 10^{42}$~erg~s$^{-1}$ in 2--10 keV and 10--40 keV, respectively. The best-fit spectrum with corresponding residuals for AGN in the isolated MCG -02-09-040 is presented in Fig.~\ref{fig2}, bottom-left panel, see also Table~\ref{tab2}.

\textbf{ESO 438-009.} As a first fit to the NuSTAR spectra we used a simple cut-off power-law model \textit{cutoffpl}. The quality of fit is very good ($\chi^{2}$/d.o.f.= 188/223) and the best-fit values of spectral index is $\Gamma = 1.73 \pm 0.06$. No additional neutral or ionized absorption except Galactic, was needed for fitting.

As the next approach, to check whether a reflection component exists in the spectrum, we replaced \textit{cutoffpl} model with the \textit{pexmon} model. This also gave an excellent fit $\chi^{2}$/d.o.f.=186/221 with photon index $\Gamma = 1.79 \pm 0.06$ and reflection parameter of $R=0.20_{-0.16}^{+0.20}$, suggesting a low ''cold'' reflection component which is consistent with the absence of obvious Fe $K_{\alpha}$ line.

The inclusion of the Swift/XRT data (0.5--8 keV energy range) clearly reveals a soft excess, which was successfully fitted ($\chi^{2}$/d.o.f.=340/351) by addition power-law component in \textit{po+cutoffpl} model. The values of power-law indices are $\Gamma_{soft}= 2.98_{-0.76}^{+0.83}$ and $\Gamma_{hard}=1.69 \pm 0.05$. We added also the Swift/BAT data in 14--150 keV energy band. A fitting of this combined data by the last model provides a good fit ($\chi^{2}$/d.o.f.=349/357) and shows a similar value of hard spectral index $\Gamma_{hard}=1.68_{-0.08}^{+0.05}$ and allowed us to obtain estimation of the cut-off energy $E_{cut} > 117$ keV. 

We also replaced the \textit{po+cutoﬀpl} model with the {pexmon} model again. This refitting is resulted in $\chi^{2}$/d.o.f. = 348/356 with $\Gamma=1.78 \pm 0.04$, reflection $R=0.25 \pm 0.16$ and cut-off energy of $E_{cut} > 115$ keV. We note that we do not need an additional soft component in the case of fit with the reflection model. 

Finally, we replaced the \textit{pexmon} model with the more physically realistic \textit{xillver} model plus primary cut-off power-law component. Here we specifically employ \textit{xillver-Ec4}, which allows for variable cut-off energy up to 1 MeV and viewing angle (inclination). We tied the power-law index and cut-off energy of \textit{xillver} to those of the cut-off power-law component. The fit statistic is again perfect, with $\chi^{2}$/d.o.f. = 347/356 (see Fig.~\ref{fig2}, upper-right panel). From this model we obtained the power-law index of $\Gamma = 1.74 \pm 0.03$; the reflection parameters yield an ionisation state of log$\xi = 3.03_{-0.27}^{+0.26}$. In this case, the cut-off energy is not be constrained by the fit.

We define a phenomenological reflection flux fraction, $R_{f}$, as the ratio between the 10--40 keV unabsorbed flux of the reflection \textit{xillver} component and power-law continuum component\footnote{Flux values were calculated using \textit{cflux} convolution model} (see e.g. \cite{Garcia2011}, \cite{Sambruna2011}, \cite{Vasylenko2018a}). However, this value cannot be compare directly with reflection fraction $R$ in \textit{pexrav} or \textit{pexmon} models, but it can be used to obtain an estimated value of reflection. This gives us reflection value $R_{f} \approx$ 0.17. Interestingly, that \textit{pexmon} is a ''cold'' reflection model, whereas \textit{xillver} is an ionized reflection, but the obtained $R_{f}$ value is comparable with the first one taking into account the error bars for $R$.

The unabsorbed luminosities are L$_{2-10~keV}=4.73_{-0.13}^{+0.10} \times 10^{42}$~erg~s$^{-1}$ and L$_{10-40~keV}=5.92_{-0.20}^{+0.16} \times 10^{42}$~erg~s$^{-1}$ in 2--10 keV and 10--40 keV, respectively (see also Table~\ref{tab2}).

\textbf{IGR J11366-6002}. As a first fit to the NuSTAR spectra we used a simple cut-off power-law model \textit{cutoffpl}. This provides a good fit with $\chi^{2}$/d.o.f.=282/272 and spectral index $\Gamma$ = 1.76 $\pm$ 0.06. Further, we replaced \textit{cutoffpl} model with a reflected model \textit{pexmon}. This model provides a very good fit $\chi^{2}$/d.o.f. = 267/271 with a softer power-law index $\Gamma = 1.97_{-0.11}^{+0.10}$ as well as a presence of moderate reflection component $R = 0.98_{-0.49}^{+0.64}$. We applied also a more physically reflection model in combination with power-law \textit{xillver+power-law} and tied the \textit{xillver} photon index to power-law photon index. This approach leads to a fit result of $\chi^{2}$/d.o.f. = 257/271. The spectral index and ionization parameter are constrained to $\Gamma = 1.76 \pm 0.05$ and $\xi = 1503_{-362}^{+638}$ erg~cm~s$^{-1}$.

The combined NuSTAR and Swift/XRT spectra showed the presence of additional partial covering neutral absorption. Fitting the data with \textit{zpcfabs*mekal+pexmon} model provides a good fit ($\chi^{2}$/d.o.f. = 335/359) with the best-fit parameters of $\Gamma = 2.00 \pm 0.05$, $R = 0.95_{-0.26}^{+0.28}$, neutral absorption $N_{H}=1.65_{-0.56}^{+0.73} \times 10^{22}$~cm$^{-2}$ and covering factor $f_{c}=0.67_{-0.07}^{+0.06}$. We note that the values of power-law index and reflection are similar to those with fitting only the NuSTAR data. If we replace the \textit{pexmon} with the \textit{xillver+power-law} model, we obtain an excellent fit to the data ($\chi^{2}$/d.o.f. = 334/358) and slightly different spectral parameters: the photon index $\Gamma = 1.78 \pm 0.05$, ionization parameter $\xi = 1364_{-277}^{+788}$ erg~cm~s$^{-1}$, neutral absorption $N_{H}=1.45_{-0.35}^{+0.58} \times 10^{22}$~cm$^{-2}$ and covering factor $f_{c}$~>~0.21.

The finally combined NuSTAR and Swift/XRT+BAT spectrum allows us to determine the cut-off energy using \textit{xillver+cutoffpl} model. In this case, we also applied \textit{xillver-Ec4}, as for previous case of ESO 438-009. We also tied the \textit{xillver} photon index to power-law photon index. We also additionally fixed inclination angle to $i=60^\circ$. This provides a good fit to the data, with $\chi^{2}$/d.o.f. = 341/364. The best-fit parameters were $\Gamma = 1.79 \pm 0.04$, 
$E_{cut} > 196$ keV, log$\xi = 3.09_{-0.08}^{+0.19}$, $N_{H}=1.37_{-0.39}^{+0.52} \times 10^{22}$~cm$^{-2}$ and only a low limit for covering factor $f_{c}$ > 0.8. The calculated reflection fraction for this fit is $R_{f} \approx$ 0.57. The cut-off energy is neither properly constrained when only cut-off power-law is used, nor when the reflection models \textit{pexmon} or \textit{xillver} are used.

We note that \cite{Masetti2008} have found that this galaxy is characterized as Compton thin Seyfert 2. These authors used optical spectroscopy, in particular, [O III] 5007 {\AA} line flux measurements. Our result is entirely consistent with this previous conclusion.

The unabsorbed luminosities are L$_{2-10~keV}=1.39_{-0.01}^{+0.02} \times 10^{42}$~erg~s$^{-1}$ and L$_{10-40~keV}=1.87 \pm {0.08} \times 10^{42}$~erg~s$^{-1}$ in 2--10 keV and 10--40 keV, respectively. The best-fit spectrum with corresponding residuals for AGN in the isolated IGR J11366-6002 is given in bottom-right panel of Fig.~\ref{fig2}, see also Table~\ref{tab2}.

\section{Discussion and conclusions}\label{sec4}
We presented the spectral analysis of four active nuclei of 2MIG isolated galaxies, namely NGC 5347, MCG -02-09-040, ESO 438-009 and IGR J11366-6002. Results for MCG -02-09-040 are presented for the first time. These galaxies are very different in terms of final model of their spectra and parameters of these models. The absence of additional column absorption characterizes NGC~5347, but MGC-02-09-040 demonstrates opposite signs of highly obscured AGN. The best-fit power-law index lies in the range of 1.6-1.8. We also were to obtain the estimation of cut-off energy for NGC 5347 and ESO 438-009, namely $E_{cut} > 63$ keV and $E_{cut} > 115$ keV.

At the same time, all four isolated AGNs have a relatively small scattering of absorption corrected luminosity in 2--10 keV, which is L$_{2-10~keV}=(0.3 \div 4.7) \times 10^{42}$~erg~s$^{-1}$. According to \cite{Lusso2012} we can calculate the bolometric luminosity, L$_{bol}$, using a factor of 10. The corresponding values are $2.9 \times 10^{42}$~erg~s$^{-1}$ for NGC 5347, $4.7 \times 10^{43}$~erg~s$^{-1}$ for ESO 438-009 and $1.4 \times 10^{43}$~erg~s$^{-1}$ for both of MCG-02-09-040 and IGR J11366-6002.

\cite{Kammoun2019} performed spectral analysis of NGC 5347 in 0.6--30 keV energy range using data from NuSTAR, Suzaku, Chandra and physical models \textit{MYTorus} (\cite{Murphy2009}) and \textit{Borus} (\cite{Balokovic2018}). They found a large equivalent width of the Fe $K_{\alpha}$ line EW = $2.3 \pm 0.3$ keV, which is comparable with our result; the inclination angle of torus is $\theta _{inc} \sim 62^{o} $ and suggested that the reprocessed emission comes from the far side of the torus. In contrast to our work, \cite{Kammoun2019} used an absorbed intrinsic emission component in spectral analysis. However, they had some difficulties with determining the value of photon index (in particular, even pegged at maximum soft limit of 2.4 for \textit{pexmon} model) and the value of line-of-sight column density too, that even reached a high limit for fitting with \textit{Borus} model and \textit{MYTorus} model in coupled mode. These authors have estimated also the mass of supermassive black hole (SMBH) as log M$_{BH}$/M$_{\odot}$ = 6.97 and ratio L$_{bol}$/L$_{Edd}$ for this galaxy from L$_{bol}=(1.65 \pm 0.33) \times 10^{43}$~erg~s$^{-1}$ $\simeq 0.014 \pm 0.005$ L$_{Edd}$. 

Following the estimation of SMBH mass for ESO 438-009 as log M$_{BH}$/M$_{\odot}$ = 7.97 (\cite{Koss2017}) and corresponding L$_{Edd} \simeq 1.2 \times 10^{46}$~erg~s$^{-1}$, we can calculate the Eddington ratio for this galaxy as L$_{bol}$/L$_{Edd} \simeq$ 0.004. To obtain this ratio for MCG-02-09-040 and IGR J11366-60020, we used method based on the 2MASS K magnitude for SMBH mass estimates described in \cite{Mushotzky2008}. Consequently, these values are equal to log M$_{BH}$/M$_{\odot}$ = 8.4 for MCG-02-09-040 and log M$_{BH}$/M$_{\odot}$ = 7.8 for IGR J11366-60020. The corresponding Eddington luminosity and ratio are L$_{Edd} = 3.16 \times 10^{47}$~erg~s$^{-1}$ and L$_{bol}$/L$_{Edd}$ = 0.0004 for MCG-02-09-040, L$_{Edd} = 7.96 \times 10^{46}$~erg~s$^{-1}$ and L$_{bol}$/L$_{Edd}$ = 0.0018 for IGR J11366-60020.

We note that our preliminary estimations of masses of SMBH in isolated AGNs (\cite{Chesnok2010}) together with result presented in this paper are in agreement with that the isolated AGNs in the Local Universe have trend to contain the lower-mass black holes relative to major population of AGNs (e.g., \cite{Baron2019}, \cite{Woo2002}).
        
All the studied objects shows a low accretion rate, from 1.5 \% of the Eddington limit to very low $\sim$ 0.04 \% that in agreement with the theoretically predicted radiative inefficient accretion flow (RIAF) (\cite{Narayan1998, Ho2009}). Concerning this result, an interesting question appears: Are the low Eddington ratios caused by intrinsic evolution of these galaxies, which did not interact with other galaxies during at least 3 Gyrs?
In the case of a very low accretion scenario such structures as the broad-line region and obscuring torus should be disappeared (\cite{Ho2008}), but the results of spectral analysis for MCG-02-09-040 and NGC 5347 are in contradiction with this point of view, at least, when existence of obscuring structure is observed. \cite{Melnyk2013} in their research with two selected samples from XMM-LSS field (X-ray galaxies and AGNs) have studied an environment effect and found that AGNs, including soft X-ray AGNs, when comparing to X-ray galaxies, prefer to be located in lower galaxy overdensities. It is in a good agreement with our estimations on the influence of choice of isolation criteria for spatial analysis and physical/morphological properties of host galaxies (\cite{Vavilova2009, Elyiv2009}). The influence of X-ray flux fluctuations to optical flux was found in several AGNs \cite{Chesnok2009} from our list. Is this phenomenon originates only in non-isolated AGNs or could be represented in isolated ones as well? The answer for this question can leads us to better understanding of AGNs structure and role of interactions with another galaxies on their structure at least in pairs. We are planning to prepare a more detailed comparative analysis of X-ray characteristics of the isolated AGNs at z < 0.1 selected from 2MIG catalogue (\cite{Chesnok2010b, Pulatova2015, Vavilova2015c, Vasylenko2015}), AMIGA project (\url{http://amiga.iaa.es/p/1-homepage.htmand}) and other databases.

\begin{appendix}
\section{SPECTRAL CHANGES OF IGC J11366-6002 FROM THE NUSTAR DATA}

In this Appendix, we discuss whether we could safely use the time-averaged spectrum in our analysis instead of several time-resolved spectra of the IGC J11366-6002. Indeed, we can expect that some spectral parameters may change due to the observed variability in the light curve of this source (see, Fig.~\ref{fig2}, bottom-right panel). 

\begin{figure}[ht!]
\epsfig{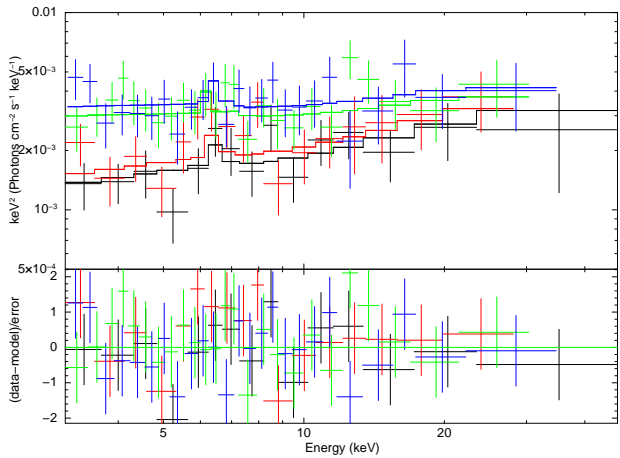}
\caption{Plot with the time-resolved NuSTAR spectra for two periods:``low part'' of light curve - red and black data points vs. ``bright part'' - green and blue data points. FMPA and FPMB detectors represented as the first and second colours respectively. \label{figA1}}
\end{figure}

We performed the addition spectral analysis of NuSTAR observations to verify this suggestion. We chose two periods: one in the middle of the observation (``bright part'') and another at the end of NuSTAR observation (``low part''). Both of parts have duration of $\sim 3500$ s. We adopted $\textit{tbabs} \ast \textit{pexmon}$ model for the fit without ``soft'' part because we use spectral data above 3 keV. Due to a low quality of spectra, we fixed the reflection parameter equal to that in the main analysis, i.e. $R \equiv 0.9$. Consequently, the following results were obtained: $\Gamma_{low} = 1.87_{-0.15}^{+0.16}$ for ``low part'' light curve, and $\Gamma_{bright} = 2.13_{-0.12}^{+0.13}$ for "bright part" light curve (Fig.~\ref{figA1}). The statistics was excellent $\chi^{2}$/d.o.f. = 55/77 for this simultaneous fit. At the same time, we obtained $\Gamma = 1.97_{-0.11}^{+0.10}$ during the main fit. Thus, we can see that all of these three power-law indexes overlap within their errors ranges. Therefore, we can safely use the time-averaged spectrum for our analysis. However, the X-ray data with better quality would be preferable for performing the detailed time-resolved spectral analysis.

We also note that values of $\Gamma_{low}$ and $\Gamma_{bright}$ evident in favor of the widely known spectral trend for AGNs with Eddington ratios above 0.01, namely, ``softer-when-brighter behavior''(e.g., \cite{Sobolewska2009}), and this behavior can be understood in terms of changes of the coronal temperature (e.g., \cite{Soldi2014}).

\end{appendix}

\section*{Acknowledgments}
This work is based on observations obtained with: the NuSTAR mission, project led by the California Institute of Technology, managed by the Jet Propulsion Laboratory and funded by NASA; the Swift satellite and it made use of data supplied by the UK Swift Science Data Centre at the University of Leicester.

This research has made use of the NuSTAR Data Analysis Software (NuSTARDAS) jointly developed by the ASI Science Data Center (ASDC, Italy) and the California Institute of Technology (USA). This research has made use both of the Palermo BAT Catalogue and database operated at INAF -- IASF Palermo and of the 70 month Swift-BAT Catalogue. data, software and/or web tools obtained from NASA High Energy Astrophysics Science Archive Research Center (HEASARC), a service of Goddard Space Flight Center and the Smithsonian Astrophysical Observatory.

The authors thank the referee for the helpful remarks, which allowed us to improve the paper.  

This work was performed in frame of the the Target Complex Program of Space Science Research of the National Academy of Sciences of Ukraine (2018--2022).

\nocite{*}
\bibliography{vasvavpul}%

\begin{landscape}
\begin{table}[bt]
\centering
\begin{threeparttable}
\caption{General information about NGC 5347, MCG -02-09-040, ESO 438-009, IGR J11366-6002 and their observations with NuSTAR and Swift instruments.\label{tab1}}
\begin{center}
\fontsize{8}{13}\selectfont
\begin{tabularx}{0.84\linewidth}{p{0.8in}|p{0.5in}p{0.5in}p{0.3in}p{0.35in}p{0.35in}p{0.3in}p{0.65in}p{0.58in}p{0.65in}p{0.35in}p{0.6in}}
\toprule
\textbf{Name} & \textbf{RA} & \textbf{Dec} & \textbf{Activ. type}\tnote{1} & \textbf{Gal. morph.}\tnote{2} & \textbf{z} & \textbf\textit{N}$_{H}$\tnote{3} & \textbf{NuSTAR \newline obs. date} & \textbf{Exp. FPMA/ \newline FPMB}\tnote{4} \newline \textbf{(ks)} & \textbf{Swift/XRT \newline obs. date} & \textbf{Exp. XRT}\tnote{5} \newline \textbf{(ks)} & \textbf{Exp. BAT}\tnote{6} \newline \textbf{(s)} \\
\midrule
\textbf{NGC 5347} & 208.3243 & 33.4908 & Sy2 & Sab & 0.0078 & 0.015 & 2015-01-16 & 47.3/ \newline 47.2 & 2009-05-07 \newline 2009-05-08 \newline 2009-05-09 \newline 2015-01-17 & 2.2 \newline 6.1 \newline 2.1 \newline 1.9 & -- \\
\textbf{MCG -02-09-040} & 51.2706 & -12.3079 & Sy2 & S0-a & 0.0149 & 0.053 & 2015-02-25 & 21.5/ \newline 21.5 & 2010-10-28 \newline 2010-11-04 \newline 2010-11-24 & 1.1 \newline 3.2 \newline 4.3 & $185.4 \times 10^{6}$ \\
\textbf{ESO 438-009} & 167.7000 & -28.5010 & Sy1 & SBab & 0.0240 & 0.052 & 2015-02-01 & 21.7/ \newline 21.6 & 2010-11-02 \newline 2010-11-03 \newline 2010-11-07 \newline 2015-02-01 & 1.2 \newline 6.4 \newline 3.3 \newline 6.5 & $7.243 \times 10^{6}$ \\
\textbf{IGR J11366-6002} & 174.1752 & -60.0518 & Sy2 & Sa & 0.0140 & 0.621 & 2014-10-29 & 21.6/ \newline 21.5 & 2007-06-22 \newline 2007-10-02 \newline 2008-02-09 \newline 2014-10-29 \newline 2014-10-30 & 4.7 \newline 5.1 \newline 3.1 \newline 1.6 \newline 4.9 & $9.525\times10^{6}$ \\
\bottomrule
\end{tabularx}
\begin{tablenotes}
        \footnotesize
        \item[1] Optical classification of Seyfert type.
        \item[2] Galaxy morphology from HYPERLEDA database.
	\item[3] Galactic absorption in units of $10^{22}$~cm$^{-2}$.
        \item[4] FPMA/FPMB observation exposures.
        \item[5] XRT in pc mode observation exposures.
        \item[6] Swift/BAT observation exposures. For ESO 438-009 and IGR J11399-6002 there are durations from Swift BAT 70-Month Hard X-ray Survey. For MCG-02-09-040 this is duration from Swift/BAT 105-Month Hard X-ray Survey.
    \end{tablenotes}
    \end{center}
    \end{threeparttable}
    \end{table}
    \end{landscape}

\begin{landscape}
\begin{table}[bt]
\centering
\begin{threeparttable}
\begin{center}
 
\caption{Best-fit spectral parameters with maximum of broad-band energy range.\label{tab2}}
\fontsize{8}{13}\selectfont
\begin{tabularx}{1.01\linewidth}{l|cccc|ccccc|ccccc}
\toprule
\textbf{Baseline model}  & \multicolumn{4}{c|}{\textbf{\tt cutoffpl}} & \multicolumn{5}{c|}{\textbf{\tt pexmon}}   & \multicolumn{5}{c}{\textbf{\tt xillver}}   \\
\midrule
  \textbf{Object name} & $\Gamma$  &  \specialcell{{\it N}$_{H}$, \\ $10^{22}$~cm$^{-2}$ }& \specialcell{{\it E}$_{cut-off}$, \\ keV} &  {$\chi^2$/d.o.f.}  &  $\Gamma$  & R  & \specialcell{{\it N}$_{H}$, \\ $10^{22}$~cm$^{-2}$ }  &  \specialcell{{\it E}$_{cut-off}$, \\ keV}  & {$\chi^2$/d.o.f.} & $\Gamma$ & log $\xi$ & \specialcell{{\it N}$_{H}$, \\ $10^{22}$~$cm^{-2}$ }  &   \specialcell{{\it E}$_{cut-off}$, \\ keV} & {$\chi^2$/d.o.f.}  \\ \hline
\textbf{NGC 5347}          & - & -  & - &  -       & 1.60$_{-0.27}^{+0.37}$ & 1(f) & - & 117$_{-54}^{+357}$  & 55/52\tnote{1}          & - & - & - & - & -              \\ 
\textbf{MCG-02-09-040}     & 1.53$_{-0.11}^{+0.12}$  & 112$_{-16}^{+20}$ &  450$_{-342}^{+peg}$ & 72/75          &  2.13$\pm$0.13 & 1(f) & 186$_{-40}^{+45}$  & - & 77/75           & - & - & - & - & -            \\ 
\textbf{ESO 438-009}       & 1.68$_{-0.08}^{+0.05}$  &  - &  234$_{-117}^{+peg}$  & 349/357        & 1.78$\pm$0.04  &  0.25$\pm$0.16 & - & 218$_{-103}^{+849}$  & 348/356         &  1.74$\pm$0.03 & 3.03$_{-0.26}^{+0.27}$  & -  & - & 347/356       \\ 
\textbf{IGR J11366-6002}   & -  & -  & - & -        & - & - & - & - & -          & 1.79$\pm$0.04  & 3.09$_{-0.08}^{+0.19}$ & 1.37$_{-0.39}^{+0.52}$ & 430$_{-234}^{+peg}$ & 341/364       \\
\bottomrule
\end{tabularx}
\begin{tablenotes}
  \footnotesize
     \item[1] C-stat
\end{tablenotes}
\end{center}
\end{threeparttable}

\vskip20pt

\begin{center}
\caption{Best-fit spectral parameters for physical torus model.\label{tab3}}
\fontsize{8}{13}\selectfont

\begin{tabular}{c|ccccc}
\toprule
\textbf{Baseline model}  &  \multicolumn{5}{c}{\tt BNTorus}    \\
\midrule
\textbf{Object name} &  $\Gamma$ & \specialcell {\it{N}$_{H}$, \\ $10^{22}$~cm$^{-2}$ }  & \specialcell{${\Theta_{incl}}$, \\ deg}  & \specialcell{ ${\Theta_{oa}}$, \\ deg}  & {$\chi^2$/d.o.f.} \\ \hline
MCG-02-09-040          &  1.63$\pm$0.11  & 104$_{-21}^{+26}$  &  41$_{-8}^{+26}$ & 37$_{-peg}^{+7}$  & 77/76           \\
\bottomrule

\end{tabular}
\end{center}
\end{table}
\end{landscape}

\end{document}